\documentclass[aps,prd,eqsecnum,nofootinbib,showpacs,twocolumn]{revtex4-1}
\usepackage{amsmath,amssymb,bm}
\usepackage[dvips]{graphicx}
\usepackage{color}

\newcommand{\be}{\begin{equation}}
\newcommand{\ee}{\end{equation}}
\newcommand{\bea}{\begin{eqnarray}}
\newcommand{\eea}{\end{eqnarray}}

\newcommand{\eq}[1]{Eq.~(\ref{eq:#1})}
\newcommand{\sect}[1]{Sec.~\ref{sec:#1}}
\newcommand{\appen}[1]{Appendix~\ref{sec:#1}}

\newcommand{\del}{\partial}
\newcommand{\eT}{\epsilon_T}
\newcommand{\ds}{d_s}

\newcommand{\bareta}{\bar{\eta}}
\newcommand{\barw}{\bar{w}}

\newcommand{\bra}{\langle}
\newcommand{\ket}{\rangle}
\newcommand{\calO}{{\cal O}}

\bmdefine{\bmk}{{\bm{k}}}
\bmdefine{\bmp}{{\bm{p}}}
\bmdefine{\bmq}{{\bm{q}}}
\bmdefine{\bmx}{{\bm{x}}}
\bmdefine{\bmPi}{{\bm{\pi}}}
\bmdefine{\bmo}{{\bm{0}}}
\bmdefine{\bmJ}{{\bm{J}}}
\bmdefine{\bmnabla}{{\bm{\nabla}}}
\bmdefine{\bmf}{{\bm{f}}}
\bmdefine{\bmv}{{\bm{v}}}
\bmdefine{\bmE}{{\bm{E}}}

\newcommand{\tilG}{\Tilde{G}}

\newcommand{\hp}{\Hat{p}}

\begin{document}

%%%   Title Page   %%%

\title{Dynamic universality class of large-$N$ gauge theories}
\author{Makoto Natsuume}
\email{makoto.natsuume@kek.jp}
\affiliation{KEK Theory Center, Institute of Particle and Nuclear Studies, 
High Energy Accelerator Research Organization,
Tsukuba, Ibaraki, 305-0801, Japan}
\author{Takashi Okamura}
\email{tokamura@kwansei.ac.jp}
\affiliation{Department of Physics, Kwansei Gakuin University,
Sanda, Hyogo, 669-1337, Japan}
\date{\today}
\begin{abstract}
In dynamic critical phenomena, singular behaviors appear not only in the order parameter but also in the other transport coefficients (due to the mode-mode coupling). However, this effect has not been observed in the AdS/CFT duality. We point out that this mode-mode coupling is suppressed by $1/N^2$ in the large-$N$ gauge theories which correspond to model H in the classification of Hohenberg and Halperin. Thus, the effect cannot be seen in the classical supergravity approximation. We illustrate this point using the example of the ${\cal N}=4$ super-Yang-Mills theory at a finite chemical potential. We also discuss the implications of this result to heavy-ion collisions.
\end{abstract}
%, 74.20.-z} %?
% KEK-TH-1424

\maketitle

%%%%%%%%%%%%%%%%%%%%%%%%%%%%%%%%%%%%%%%%
\section{Introduction} \label{sec:intro}
%%%%%%%%%%%%%%%%%%%%%%%%%%%%%%%%%%%%%%%%

In the second-order phase transitions, the correlation length $\xi$ diverges, and as a result various thermodynamic quantities diverge \cite{textbook}. In the dynamic case (dynamic critical phenomena), the relaxation time of the order parameter also diverges, which is known as the critical slowing down. 
%This appears as a singular behavior in the transport coefficients related to the order parameter. 
In the context of the AdS/CFT duality \cite{Maldacena:1997re,Witten:1998qj,Witten:1998zw,Gubser:1998bc}, the dynamic critical phenomena have been studied in Refs.~\cite{Maeda:2008hn,Maeda:2009wv,Buchel:2009mf,Buchel:2010gd,Natsuume:2010vb,Buchel:2010ys}. 

In the dynamic critical phenomena, singular behaviors are not limited only to the quantities related to the order parameter. In general, if the system has conserved quantities (such as the energy-momentum tensor $T^{\mu\nu}$), the dynamic universality class and the dynamic critical exponent change due to the coupling between conserved charges and the order parameter (mode-mode coupling). As a result, one has singular behaviors in the other transport coefficients (such as the shear viscosity $\bareta$.%
\footnote{In this paper, we use $\bareta$ for the shear viscosity to avoid confusion with a static critical exponent $\eta$.})

To be more specific, let us specify a class of theories we consider. Below we consider a class of gauge theories with local $SU(N_c) \times$ global $U(1)$ symmetry. So, the gauge theory has a conserved charge associated with the global $U(1)$ symmetry, and we consider the case where the charge is an order parameter of a second-order phase transition. As we discuss below, this class includes the ${\cal N}=4$ super-Yang-Mills (SYM) theory at a finite chemical potential and QCD. According to the classification of Hohenberg and Halperin \cite{hohenberg_halperin} (see \sect{classification} for the classification), the dynamic universality class of these theories is model H. Model H has a conserved order parameter with additional conserved quantities such as $T_{\mu\nu}$. Model H is a typical universality class for a field theory since a field theory has the energy-momentum tensor. This class shows singular behaviors in the charge conductivity $\lambda$ and the shear viscosity $\bareta$ due to the mode-mode coupling (\sect{H}).

We point out that this mode-mode coupling is suppressed by $1/N_c^2$ for large-$N_c$ gauge theories. Namely, this effect cannot be seen in the large-$N_c$ limit, {\it i.e.}, the gauge coupling $g_{\rm YM} \rightarrow 0, N_c \rightarrow \infty$ with a fixed large $\lambda:=g_{\rm YM}^2 N_c$. From the gravity point of view, this mode-mode coupling effect cannot be seen in the classical supergravity approximation, and it should emerge from a one-loop computation. We illustrate this point in detail using the example of the ${\cal N}=4$ SYM. 
%super-Yang-Mills (SYM) theory at a finite chemical potential.

The ${\cal N}=4$ SYM can undergo a second-order phase transition at a finite chemical potential. The gravity dual of the system is known as the five-dimensional R-charged black hole (which is also known as ``spinning" D3-brane or STU black hole) \cite{Behrndt:1998jd,Kraus:1998hv,Cvetic:1999xp}. Its dynamic critical phenomenon has been studied in Refs.~\cite{Maeda:2009wv,Buchel:2010gd}. 
In this system, the charge density shows a singular behavior, and the dynamic universality class is model B. 
%. As a result, the dynamic universality class is model B in the classification of Hohenberg and Halperin \cite{hohenberg_halperin}. (See \sect{classification} for the classification.) 
For model B, only the order parameter shows a singular behavior (\sect{A&B}), but the system obviously has 
%the other conserved quantities 
$T^{\mu\nu}$. Thus, one expects that the universality class becomes model H, but this is not the case in the AdS/CFT duality (see \sect{known_facts}). This is because the leading-order AdS/CFT computation cannot see the mode-mode coupling effect. 

In this paper, we mostly focus on the ${\cal N}=4$ plasma, but our argument is generic and is independent of the details of gauge theories. In \sect{heavy_ion}, we discuss the implications to QCD. Son and Stephanov have argued that QCD belongs to model H at the critical end point \cite{Son:2004iv}. Since the model H effect is suppressed by $1/N_c^2$, we discuss the implications to the heavy-ion collisions.

%%%%%%%%%%%%%%%%%%%%%%%%%%%%%%%%%%%%%%%%
\section{Dynamic universality class} %\label{sec:intro}
%%%%%%%%%%%%%%%%%%%%%%%%%%%%%%%%%%%%%%%%

%%------------------
\subsection{Hydrodynamic variables}
%%------------------

In the study of critical phenomena, one begins to identify the order parameter and to study the statics. However, in order to determine the dynamic universality class, the full set of macroscopic variables  (hydrodynamic variables) may become important as we will see below. 

%The first step to determine the dynamic universality class is to identify macroscopic variables (hydrodynamic variables). 
Typical macroscopic variables are
\begin{enumerate}
\item conserved charges such as charge, energy, and momentum densities 
\item Nambu-Goldstone modes (if there is continuous symmetry breaking)
\item order parameter (which arises via critical slowing down in a second-order phase transition)
\end{enumerate}
These modes may not be all independent: some conserved charge may become an order parameter.
Also, not all components of $J^\mu$ and $T^{\mu\nu}$ correspond to hydrodynamic variables. Only conserved charges are hydrodynamic variables: they are guaranteed to survive in the hydrodynamic limit $\omega(q)\rightarrow 0$ as $q \rightarrow 0$ because of conservation laws. To isolate hydrodynamic variables, it is useful to carry out the tensor decomposition of $J^\mu$ and $T^{\mu\nu}$. For example, $J^\mu$ can be decomposed as the longitudinal diffusive mode and the transverse modes. The charge density $\rho$ appears in the diffusive mode. 
%Similarly, for $T^{\mu\nu}$, the energy and momentum densities appear in the longitudinal sound mode and the transverse shear mode, respectively. 
Similarly, for $T^{\mu\nu}$, the energy and longitudinal momentum density appear in the sound modes and the transverse momentum densities appear in the shear modes. 
%\marginpar{causal?}

%%------------------
\subsection{Static universality class} \label{sec:statics}
%%------------------

In static critical phenomena, various static quantities associated with the order parameter diverge at the critical point. The divergences are parametrized by static critical exponents. These exponents define a static universality class.

Traditionally, there are six static critical exponents. For ferromagnets, they are defined as follows:
the specific heat: $C_H \propto |\eT|^{-\alpha}$, 
the spontaneous magnetization: $m \propto |\eT|^{\beta}~(T<T_c)$, 
the magnetic susceptibility: $\chi \propto |\eT|^{-\gamma}$,
the critical isotherm: $m \propto |h|^{1/\delta}  (T=T_c)$,
the correlation function $(T \neq T_c)$: $G(r) \propto e^{-r/\xi}$, 
the correlation function $(T=T_c)$: $G(r) \propto r^{-d_s+2-\eta}$, and 
the correlation length: $\xi  \propto |\eT|^{-\nu}$.
Here, $m$ is the magnetization, $h$ is the external magnetic field, $\eT:= (T-T_c)/T_c$, and $d_s$ denotes the number of {\it spatial} dimensions.
%Definition of static critical exponents:
%\be
%\begin{array}{rcll}
%C_H
%C &\propto& |\eT|^{-\alpha}~, & \\
%m &\propto& |\eT|^{\beta} & (T<T_c)~, \\
%\chi_T
%\chi &\propto& |\eT|^{-\gamma}~, & \\
%m & \propto& |h|^{1/\delta} & (T=T_c)~, \\
%G(r)  & \propto& e^{-r/\xi} & (T \neq T_c)~, \\
%        & \propto& r^{-\ds+2-\eta} & (T=T_c)~, \\
%\xi  &\propto& |\eT|^{-\nu}~. &
%\end{array}
%\ee
For the Ginzburg-Landau theory, 
\be
(\alpha, \beta, \gamma, \delta, \nu, \eta) = \left(0, \frac{1}{2}, 1, 3,  \frac{1}{2}, 0\right)~.
%_{\rm GL}~.
\label{eq:static_exponent_GL}
\ee 
%In this case,
%\be
%T-T_c \quad\leftrightarrow\quad \xi^{-2}
%\ee
These exponents are not all independent, but they satisfy static scaling relations:
\be
\alpha + 2\beta+\gamma =2~, %\quad
\gamma = \beta(\delta-1)~, %\quad
\gamma = \nu (2-\eta)~, %\quad
2-\alpha = \nu \ds~
\label{eq:scaling}
\ee
except the last hyperscaling relation, which often fails (\sect{N=4_static_universality}). The scaling relations suggest that there is some structure behind them, which is known as the scaling law.

%%------------------
\subsection{Dynamic universality class} \label{sec:classification}
%%------------------

%In static critical phenomena, various thermodynamic quantities associated with the order parameter diverge. 
In dynamic critical phenomena, the relaxation time $\tau$ of the order parameter also diverges as $\tau \propto \xi^z$, which is known as the critical slowing down. The exponent $z$ is the dynamic critical exponent. The effect of slow dynamics is not limited to the order parameter. The other transport coefficients also have singular behaviors in general due to couplings with the order parameter.

The details of the dynamic exponent depend on dynamic universality classes. The dynamic universality classes were classified by Hohenberg and Halperin \cite{hohenberg_halperin}: they are known as model A, B, C, (D, E), F, G, H, and J.%
\footnote{Model D and E are written in parentheses since model D reduces to model B and model E is a special case of model F with an additional discrete symmetry.}
We review model A and B in the next subsection and review model H in \sect{H}. The classification is based on
\begin{enumerate}
\item whether the order parameter is conserved or not,
\item whether there are the other conserved charges,
\item whether there exists couplings between these modes. %(nonzero Poisson brackets).
\end{enumerate}
The dynamic universality class depends on additional properties of the system which do not affect the static universality class. In particular, conservation laws play an important role to determine the dynamic universality class. A conservation law forces the relaxation to proceed more slowly. As a consequence, even if two systems belong to the same static universality class, they may not belong to the same dynamic universality class. 

%%------------------
\subsection{Models with order parameter only}\label{sec:A&B}
%%------------------

Let us start with the simplest systems where only the order parameter matters near the critical point. These are model A and B.

For model A, the order parameter is not conserved (such as the uniaxial antiferromagnet). In this case, the equation of motion for the order parameter takes the form ($\omega$: frequency)
\be
\omega = -i \Gamma_\calO~.
\label{eq:model_A}
\ee 
The relaxation rate $\Gamma_\calO \rightarrow 0$ at the critical point because the Hamiltonian has a flat potential there. Thus, the relaxation time $\tau \sim 1/\Gamma_\calO$ diverges at the critical point. For the Ginzburg-Landau Hamiltonian, $\Gamma_\calO \propto (T-T_c) \propto \xi^{-2}$, so $\tau \propto \xi^{2}$, namely $z=2$.%
\footnote{More generally, when the system has the anomalous exponent $\eta$, $z=2-\eta$ (van Hove theory) or $z = 2+ c \eta $ where $c = 0.7261(1-1.69 \epsilon+\cdots)$ (in the $\epsilon$-expansion).
%$z=2+c\eta$. The constant $c$ depends on the technique one employs.
}

For model B, a conserved charge, say $\rho$, is the order parameter (such as the uniaxial ferromagnet). A conserved charge satisfies the diffusion equation ($q$: wave number)
\be
\omega = -iDq^2~,
\label{eq:diffusion}
\ee
where $D$ is the diffusion constant. The diffusion constant $D$ is related to the susceptibility $\chi := \del \rho/\del \mu$ ($\mu$: chemical potential for the charge) and the conductivity $\lambda$ as 
\be
D=\lambda \chi^{-1}~.
\label{eq:diffusion_vs_conductivity}
\ee
This equation has the following interpretation. On one hand, the charge fluctuation $\rho$ diffuses by the charge diffusion $\bmJ = - D\bmnabla\rho$. The same process can be regarded as the charge conduction under an electric field $\bmE$:
\be
\bmJ = - D \bmnabla \rho 
= - D \left( \frac{\del\rho}{\del\mu} \right) \bmnabla \mu
= \lambda \bmE~.
\label{eq:diffusion_as_conduction}
\ee
Namely, the charge fluctuation $\rho(\bmx)$ generates a chemical potential $\mu(\bmx)$ which gives $\bmE=-\bmnabla\mu$. From \eq{diffusion_as_conduction}, one gets \eq{diffusion_vs_conductivity}.

For model B, $\lambda$ remains finite at the critical point. Because $\chi$ diverges at the critical point, $D \rightarrow 0$ and $\tau \sim - ({\rm Im}\, \omega)^{-1} \sim (Dq^2)^{-1}$ diverges. 
%The critical exponent is extracted as follows. The dispersion relation (\ref{eq:diffusion}) applies in the hydrodynamic regime ($q\xi \ll 1$). In the critical regime  ($q\xi \gg 1$), higher powers of $q$ in the dispersion relation are no longer negligible. In this regime, $\omega \sim q^z$. These two behaviors should match smoothly at $q\xi \sim 1$. 
In order to extract the dynamic critical exponent, set $q\xi \sim 1$ in the dispersion relation (\ref{eq:diffusion}), and determine $z$ from $\omega \sim \xi^{-z}$. 
%extrapolate the dispersion relation (\ref{eq:diffusion}) in the regime $q\xi \sim 1$, 
Using the definition of static exponents $\gamma, \nu$ and a scaling relation (\ref{eq:scaling}), one gets $\chi \sim \xi^{\gamma/\nu} \sim \xi^{2-\eta}$. Thus, 
\bea
\omega &=& - i \lambda \chi^{-1} q^2 
\sim \xi^{-2+\eta} q^2 \sim \xi^{-(4-\eta)} (q\xi)^2 \\
&\sim& \xi^{-(4-\eta)} \qquad (\mbox{for } q\xi \sim 1)~,
\label{eq:diffusion_extrapolate}
\eea
so
\be
z = 4 - \eta~.
\ee

The logic behind setting $q\xi \sim 1$ (scaling form) is as follows. We use hydrodynamic dispersion relations such as  \eq{diffusion}. It applies in the hydrodynamic regime ($q\xi \ll 1$) and takes into account the only lowest power of $q$. However, in the critical regime  ($q\xi \gg 1$), higher powers of $q$ in the dispersion relation are no longer negligible. In this regime, write $\omega \sim q^z$. (The correlation length should be dropped out in the relation at the critical point.) These two behaviors should match smoothly at $q\xi \sim 1$. The extrapolation from the hydrodynamic regime gives a relation such as \eq{diffusion_extrapolate}, and the extrapolation from the critical regime gives $\omega \sim \xi^{-z}$, so the comparison gives $z$. It is in this sense that we write $\tau \sim \xi^z$. Although we ignored this issue in the discussion of the model A exponent, a similar remark applies there as well.

As we discuss below, an example of model B is the ${\cal N}=4$ SYM in the large-$N_c$ limit.

%%%%%%%%%%%%%%%%%%%%%%%%%%%%%%%%%%%%%%%%
\section{Critical behavior of large-$N_c$ gauge theories} %\label{sec:intro}
%%%%%%%%%%%%%%%%%%%%%%%%%%%%%%%%%%%%%%%%

%%------------------
\subsection{Critical behavior of the ${\cal N}=4$ plasma: known facts} \label{sec:known_facts}
%%------------------

The ${\cal N}=4$ SYM has the $SU(4)$ R-symmetry, which is rank 3, so one can add at most 3 independent chemical potentials. We focus on the case of a single chemical potential. The theory is dual to the five-dimensional R-charged black hole, which is a solution of gravity coupled with a $U(1)$ gauge field and a scalar field. 

Let us summarize the known facts about the critical behavior of the ${\cal N}=4$ SYM in the large-$N_c$ limit derived from the R-charged black hole. For further details, see \appen{R-charged} and original references:
\begin{enumerate}

\item
The R-charge density has a singular behavior \cite{Gubser:1998jb,Cai:1998ji,Cvetic:1999rb}: the R-charge susceptibility 
\be
   \chi
  := \left( \frac{\partial \rho}{\partial \mu} \right)_{T}
  = \frac{N_c^2 T_0^2}{8}~
     \frac{2 + 5 \kappa - \kappa^2}{2 - \kappa}~
\label{eq:R-charged_susceptibility}
\ee
%$\chi_\rho := \del \rho/\del \mu$ 
diverges at $\kappa=2$, where $\kappa$ is a parameter related to the R-charge density [see \eq{charge_density}]. $T$ is the temperature which is related to $T_0$, the temperature for $\kappa=0$ [see \eq{temperature}]. So, $\kappa=2$ is the critical point, and the R-charge density is the order parameter.

\item 
The R-charge conductivity remains finite at the critical point \cite{Maeda:2008hn}: 
\be
  \lambda
%  = - \lim_{\omega \to 0}~
%      \frac{\Im\left[~G^{(R)}_{xx}(\omega, q = 0)~\right]}
%           {\omega}
  = \frac{ (\kappa + 2)^2 N_c^2 T_0 }
         { 64 \pi \sqrt{1 + \kappa} }~,
\label{eq:R-charged_conductivity}
\ee
and the diffusion constant is given by 
\be
D = \frac{1}{2 \pi T}\frac{ (1 + \kappa/2)^3 }{1 + \kappa}
     \frac{2 - \kappa}{2 + 5 \kappa - \kappa^2}
\label{eq:R-charged_diffusion}
\ee
from Eqs.~(\ref{eq:diffusion_vs_conductivity}) and (\ref{eq:R-charged_susceptibility}). This implies that the critical slowing down indeed occurs in the AdS/CFT duality, {\it i.e.}, $D\rightarrow0$. It has been shown that $z = 4-\eta$ with $\eta=0$ consistent with the model B prediction \cite{Buchel:2010gd}. 

\item The shear viscosity remains finite at the critical point:
\be
\bareta = \frac{\pi\,N_c^2\,T_0^3}{8}~\sqrt{1 + \kappa}~.
\label{eq:R-charged_viscosity}
\ee
This is related to the universality of the shear viscosity $\bareta/s=1/(4\pi)$ \cite{Mas:2006dy,Son:2006em,Saremi:2006ep,Maeda:2006by}. Since the entropy density is the first derivative of the free energy, it is continuous across the second-order phase transition. Thus, the universality implies that $\bareta$ is also continuous across the phase transition. 

\end{enumerate}
These results suggest that the ${\cal N}=4$ SYM belongs to model B rather than model H below in the large-$N_c$ limit.  However, the system obviously has the conserved quantity $T^{\mu\nu}$. In such a case, one expects that the universality class becomes model H. We try to resolve this puzzle in this paper. But first we need some basics of model H.

%%------------------
\subsection{Model H}\label{sec:H}
%%------------------

%So far we ignored the energy-momentum tensor, namely we ignored the fluid motion. Including this effect modifies the dynamic universality class and the value of the dynamic critical exponent. 

The energy-momentum tensor affects the dynamic universality class because it introduces the notion of the fluid motion which we ignored in \sect{A&B}. Including this effect modifies the dynamic universality class and the value of the dynamic critical exponent. 

The energy-momentum tensor contains the shear modes and the sound modes associated with the momentum densities and the energy density. The scaling form argument in \sect{A&B} suggests that the sound modes $\omega \sim c_s q \propto 1/\xi$ have a weak singularity compared with the shear modes $\omega \sim -i D_{\bar{\eta}} q^2 \propto 1/\xi^2 $ [$c_s$: speed of sound, $D_{\bar{\eta}} := \bareta/(\epsilon+P)$]. So, one needs to take only the shear modes into account.%
\footnote{This intuitive argument ignores the couplings among hydrodynamic modes. In \appen{H_mode_coupling}, we explicitly show that the effect of the sound modes to the order parameter is negligible for model H. On the other hand, the slow dynamics of the order parameter can affect the sound modes. As a consequence, one has singular behaviors in the speed of sound and the sound attenuation (bulk viscosity) \cite{kawasaki_bulk,onuki_bulk}. We will not consider this issue in this paper. 
%In order to justify it, one should check if coupling terms are irrelevant in the renormalization group sense. For model H, it turns out that the shear mode coupling to the charge diffusive mode is relevant for $\ds<4$ whereas the sound mode coupling to the other modes are irrelevant \cite{vasilev}. Alternatively, ... 
} 
(Here, we assume that the speed of sound and the shear viscosity either remain constant or have only a weak singularity at the critical point.)

Thus, the relevant hydrodynamic modes consist of the charge diffusive mode and the shear modes. The shear modes contain a new transport coefficient, the shear viscosity $\bareta$. These modes are known to describe model H dynamics. Here, we summarize the main results for model H and its basic physics. For further details, see Ref.~\cite{hohenberg_halperin} and references therein. 

For model H, both $\lambda$ and $\bareta$ become singular at the critical point:
\bea
D &=& \lambda \chi^{-1} \sim \xi^{x_\lambda} \chi^{-1}~, 
\label{eq:x_lambda-def} \\
\bareta &=& \xi^{x_\eta}~.
\label{eq:x_eta-def}
\eea
These new dynamic exponents satisfy the following expression: 
\be
x_\lambda + x_\eta = 4 - \ds - \eta~.
\label{eq:H_constraint}
\ee
For $\ds=3$, mode-mode coupling computations yield (assuming $\eta = 0$)
\be
x_\lambda = 0.946, \quad x_\eta = 0.054~.
\label{eq:H_results}
\ee
% v2.1
Namely, $x_\eta$ is a small number. The anomalous exponent $\eta$ is also a small number: $\eta \sim 0$ for QCD and $\eta=0$ for the ${\cal N}=4$ SYM in the large-$N_c$ limit. (See also \sect{1/N_counting} for a related discussion.) Because of the constraint (\ref{eq:H_constraint}), we eliminate $x_\lambda$ below in favor of $x_\eta$ and $\eta$ which are small numbers. While we retain these small exponents, one may ignore them for a rough estimate. 
% v2
%Hence, one may ignore $x_\eta$. One may also ignore $\eta$ since $\eta \sim 0$ for QCD and $\eta=0$ for the ${\cal N}=4$ SYM in the large-$N_c$ limit. (See also \sect{N=4_static_universality} for a related discussion.) For simplicity, we will omit them in some expressions below. These are equalities with $\approx$ which denotes equalities up to small $x_\eta$ or $\eta$.
% v1
%We henceforth ignore $\eta$ since it often vanishes. (This is the case for the R-charged black hole as well.) 

In order to know the values of $x_\lambda$ and $x_\eta$ separately, one needs mode-mode coupling computations or renormalization group analysis. However, the constraint (\ref{eq:H_constraint}) can be understood from an intuitive argument which we will see below. But first let us consider the implications of the above result: 

\begin{enumerate}

\item 
For $\ds=3$, $x_\lambda + x_\eta \sim 1$, so there is no model B limit, {\it i.e.}, one cannot obtain $x_\lambda = x_\eta = 0$. We will see the reason below. 

\item 
The singular behavior in conductivity occurs for $\ds<4$. This is related to the fact that the shear mode coupling to the charge diffusive mode is relevant in the renormalization group sense for $\ds<4$.

\item 
% v2.1
From Eqs.~(\ref{eq:x_lambda-def}) and (\ref{eq:H_constraint}), one gets
\be
D_H \sim % \xi^{x_\lambda - 2 + \eta}
\xi^{2 - \ds - x_\eta}~.
\ee
For $\ds=3$ and $x_\eta \sim 0$, 
\be
D \sim \left\{
	\begin{array}{ll}
	\xi^{-2} & \mbox{(Model B)} \\
	\xi^{-1} & \mbox{(Model H)}
	\end{array}
\right.
\ee
Namely, the diffusion constant still vanishes at the critical point but with a different power of $\xi$. The scaling argument in \sect{A&B} determines the dynamic critical exponent $z$ as 
\be
z %= 4 - \eta -x_\lambda 
= d_s + x_\eta
\ee
since $\omega = - iDq^2 \propto \xi^{2-\ds-x_\eta}q^2 =(q\xi)^2/\xi^{\ds+x_\eta}$. For $\ds=3$ and $x_\eta \sim 0$, $z \sim 3$.

\end{enumerate}

Model H does not have a model B limit. This is because the mechanism of conduction differs for these models. This is understood from an intuitive argument below. The argument also shows why the constraint (\ref{eq:H_constraint}) holds. 

For model B, the conduction comes from the charge diffusion as we saw in \sect{A&B}. However, for model H, there is an extra contribution to the conductivity from convection of fluids. The conductivity from the convection is much larger than the conductivity from the charge diffusion of model B (in the sense that the former diverges as $\xi\rightarrow \infty$). 

We introduced the external electric field $\bmE$ in \sect{A&B}. Under $\bmE$, a charged fluid will experience a mechanical force $\bmf_{\rm appl}=\rho \bmE$. The fluid will accelerate to velocity $\bmv$ at which viscous drag $\bmf_{\rm visc}$ balances the applied force $\bmf_{\rm visc}+\bmf_{\rm appl}=0$. (In this subsection, we assume $\bar{\rho}=0$, where \={ } denotes an equilibrium value. Thus, $\rho=\delta\rho$, where $\delta$ denotes the deviation from the equilibrium. Similarly, $\bmv=\delta\bmv$.) For a chunk of fluid with typical linear dimension $L$, 
\be
\bmf_{\rm visc} \sim - \bareta \bmv L^{\ds-2}~, \qquad
\bmf_{\rm appl} \sim \rho \bmE L^{\ds}~.
\ee
Then, the induced current $\bmJ$ is given by
\be
\bmJ = \rho \bmv \sim \frac{\rho^2}{\bareta} L^2 \bmE~.
\label{eq:H_current}
\ee
Thus, the conductivity $\lambda_{\mbox{\scriptsize H}}$ by convection is given by
$\lambda_{\mbox{\scriptsize H}} \sim (\rho^2/\bareta) L^2$,
%\be
%
%\lambda_{\mbox{\scriptsize H}} \sim \frac{\rho^2}{\bareta} L^2~,
%\label{eq:H_lambda1}
%
%\ee
which diverges with $L$. The divergence is cut off at the scale $L \sim \xi$ because the fluctuation 
%\be
%
$\bra \rho^2 \ket = T \chi/L^{\ds}$
%\label{eq:equipartition}
%
%\ee
is correlated at most this scale. 
%\footnote{It is important to note that $\rho^2$ in \eq{H_current} really means $\bra \rho^2 \ket$ not $\bra \rho \ket^2$. This is justified from the mode-mode coupling computation in \appen{H_mode_coupling}. } 
Therefore, 
\be
\lambda_{\mbox{\scriptsize H}} 
\sim \frac{\chi}{\bareta} \xi^{2-\ds} ~,
\label{eq:H_lambda}
\ee
or
\be
\bareta \lambda_{\mbox{\scriptsize H}} 
%\sim \bra \rho^2 \ket \xi^2
\sim \chi \xi^{2-\ds} 
\sim \xi^{4-\ds-\eta}~,
\ee
which coincides with \eq{H_constraint}. In \eq{H_current}, the current really means $\bmJ=\rho \bmv = \delta\rho\delta \bmv$, so this effect is nonlinear in deviations.

The above intuitive argument can be justified from the mode-mode coupling computation for $\eta=0$ which we reproduce in \appen{H_mode_coupling}. The computation also fixes the overall coefficient of \eq{H_lambda}. For $\ds=3$,
\be
\lambda_{\rm H} = \frac{T}{6\pi \bareta} \chi \frac{1}{\xi}~, 
\quad {\rm or} \quad
D_{\rm H} = \frac{T}{6\pi \bareta} \frac{1}{\xi}~.
\label{eq:H_mode_coupling}
\ee

%%------------------
\subsection{Large-$N_c$ counting}\label{sec:1/N_counting}
%%------------------

Generically, the conductivity has a model H (convective) contribution as well as a model B (diffusive) contribution. The AdS/CFT result (\ref{eq:R-charged_diffusion}) is the model B contribution, and it scales as $O(N_c^2)$. From the boundary point of view, this is because there are $O(N_c^2)$ degrees of freedom which carry R-charges. 

From the bulk point of view, this comes from the fact that the bulk action is proportional to $1/(16 \pi G_5)$. Namely, 
\begin{enumerate}
\item The conductivity can be derived from the two-point correlator $\bra J^iJ^j \ket$ using a Kubo formula. 
\item According to the GKP (Gubser-Klebanov-Polyakov)-Witten relation \cite{Witten:1998qj,Gubser:1998bc}, the correlator is evaluated from the on-shell bulk action. 
\item The boundary current $J^\mu$ couples to the bulk Maxwell field $A_M$. So, one needs to evaluate the on-shell bulk Maxwell action. 
\item The bulk Maxwell action is proportional to $1/(16 \pi G_5)$ and so is the correlator.%
\footnote{More generally, if the bulk Maxwell action takes the canonical form
\be
\frac{1}{g^2} \int d^5x \sqrt{-g} F_{MN}^2~,
\ee
$\chi =O(g^{-2})$, and \eq{H_1/N} changes as $ \lambda_{\mbox{\scriptsize H}} \sim O(1/(g^2 N_c^2)) \times \xi^{4-\ds}$.
} 
According to the standard AdS/CFT dictionary, $L^2/(16 \pi G_5) = N_c^2/(8\pi^2)$.
\end{enumerate}

On the other hand, the model H contribution (\ref{eq:H_lambda}) scales as $O(1)$ as $N_c \rightarrow \infty$. This is because various quantities in \eq{H_lambda} scale as follows [Eqs.~(\ref{eq:R-charged_susceptibility}) and (\ref{eq:R-charged_viscosity})]:
\be
\chi, \bareta = O(N_c^2)~,
\quad
\xi = O(1)~.
\ee
$\chi$ and $\bareta$ scale as $O(N_c^2)$ from similar reasons as $\lambda_{\rm B}$. The scaling of $\xi$ can be understood as follows. One way to derive the correlation length $\xi$ is to use the quasinormal mode method. In this case, one solves the bulk Maxwell equation to find a pole. At tree level, the Newton constant $G_5$ appears only as the overall coefficient of the bulk action, so the bulk Maxwell equation does not contain $G_5$, and $\xi$ cannot depend on $N_c$. Alternatively, one can derive $\xi$ from the static correlator $\bra \rho\rho \ket$. The correlator has a structure $\bra \rho\rho \ket \sim 1/(1+q^2\xi^2)$, where the quasinormal mode appears as a pole. Although the correlator itself is proportional to $N_c^2$, the pole is not influenced by $N_c$. 

Therefore, we conclude that the model H behavior is a subleading effect at large-$N_c$: 
\bea
&&\lambda_{\mbox{\scriptsize B}} \sim O(N_c^2) \times \xi^0~,\\
&&\lambda_{\mbox{\scriptsize H}} \sim O(1) \times \xi^{4-\ds-\eta-x_\eta}~.
\label{eq:H_1/N}
\eea
The essential reason for the $1/N_c^2$-suppression is the presence of the shear viscosity $\bareta$ in \eq{H_mode_coupling}, which is $O(N_c^2)$. For model H, the effect of the convection is more effective than the charge diffusion since the former diverges as $\xi \rightarrow \infty$. However, in large-$N_c$ gauge theories, the shear viscosity is large%
\footnote{One often says that the quark-gluon plasma and gauge theories in the large-$N_c$ limit has a very small viscosity, but it really means that $\bareta/s$ or the shear viscosity for a degree of freedom is small. The shear viscosity itself is large due to a large number of degrees of freedom.}
and $O(N_c^2)$ so that the convection turns out not to be very effective. 
%This is more clearly seen in terms of the diffusion constant.

% v2
%The anomalous exponent $\eta$ vanishes for the ${\cal N}=4$ SYM in the large-$N_c$ limit. However, static exponents are also modified by $1/N_c^2$-corrections (\sect{N=4_static_universality}). Since $\chi \propto \xi^{2-\eta}$, the corrections affect $\chi$ and $D=\lambda\chi^{-1}$, but the corrections affect neither $\lambda$ nor the ratios $\lambda_{\rm H}/\lambda_{\rm B}$ and $D_{\rm H}/D_{\rm B}$.

% v2.1
The anomalous exponent $\eta$ vanishes for the ${\cal N}=4$ SYM in the large-$N_c$ limit. However, static exponents are also modified by $1/N_c^2$-corrections (\sect{N=4_static_universality}). As seen from \eq{H_1/N}, this effect just changes the power of $\xi$ slightly and does not affect our main result. For simplicity, we ignore $\eta$ and $x_\eta$ in the following discussion.

The R-charged black holes have been constructed for $\ds \neq 3$. When one realizes the black holes using simple brane systems, black holes with $\ds =2, 3$ and 5 are particularly important (corresponding to the M2, D3, and M5-branes, respectively). In the mode-mode coupling theory, the shear mode coupling to the charge density is relevant for $\ds<4$ if fields have the canonical scaling dimensions. Thus, the singular behavior of $\lambda$ may be invisible for the $\ds=5$ case even if one takes the $1/N_c^2$-effect into account.

%%------------------
\subsection{Mode-mode coupling estimate of $1/N_c^2$-effect}
%%------------------

In the AdS/CFT duality, $G_5 \propto 1/N_c^2$. Because a supergravity 1-loop adds a weight $G_5$, the $1/N_c^2$-effect is translated into a 1-loop supergravity effect. 
Let us estimate the model H conductivity for the ${\cal N}=4$ plasma using the result of mode-mode coupling theory, which is a prediction for 1-loop supergravity.

The mode-mode coupling result is given in \eq{H_mode_coupling}. Reference~\cite{Buchel:2010gd} estimated the correlation length numerically. The correlator 
\be
\tilG(q) = \frac{T \chi}{1+(q\xi)^2}
\ee
has the pole at $q^2=-\xi^{-2}$, where
\be
\xi^{-2} \sim 0.12 (2-\kappa) (2\pi T_c)^2+ O\left((\kappa-2)^2 \right)~.
\label{eq:buchel_correlation}
\ee
Using Eqs.~(\ref{eq:H_mode_coupling}) and (\ref{eq:buchel_correlation}), one gets
\be
D_{\rm H} \sim \frac{32}{27\pi^2}\frac{1}{N_c^2T_c^2}\frac{1}{\xi}
\sim \frac{64}{27\pi}\sqrt{0.12} \frac{1}{N_c^2T_c} (2-\kappa)^{1/2}
\label{eq:R-charged_diffusion_H}
\ee
near the critical point $\kappa=2$. The corresponding conductivity is
\be
\lambda_{\rm H} \sim \frac{32}{9} 0.12\, T_c^2 \xi
\sim \frac{16}{9\pi}\sqrt{0.12}\, T_c \frac{1}{(2-\kappa)^{1/2}}~.
\label{eq:R-charged_conductivity_H}
\ee
This should be contrasted with model B results:
\bea
&& D_{\rm B} \sim \frac{1}{6\pi T_c}(2-\kappa) 
\sim \frac{1}{24\pi^3}\frac{1}{0.12 T_c^3}\frac{1}{\xi^2}~, 
\label{eq:R-charged_diffusion_B} \\
&& \lambda_{\rm B} \sim \frac{1}{8\pi}N_c^2 T_c~.
\eea

%%%%%%%%%%%%%%%%%%%%%%%%%%%%%%%%%%%%%%%%
\section{Discussion} %\label{sec:intro}
%%%%%%%%%%%%%%%%%%%%%%%%%%%%%%%%%%%%%%%%

%%------------------
\subsection{$1/N_c$-effects in the AdS/CFT duality}
%%------------------

In the AdS/CFT duality, there are various phenomena which are never visible in the large-$N_c$ limit. Examples include ``long-time power-law tails" in the relaxation phenomena \cite{Kovtun:2003vj}, the deviation from the mean-field behavior (\sect{N=4_static_universality}), and the symmetry restoration for low-dimensional symmetry breaking (due to the Coleman-Mermin-Wagner theorem). We have seen that mode-mode coupling in the dynamic critical phenomena is another example of $1/N_c$-effects. Our result indicates that the dynamic universality classes of large-$N_c$ gauge theories are simpler than the usual condensed-matter systems. %Our result indicates that the dynamic universality classes of large-$N_c$ gauge theories are not as rich as the usual condensed-matter systems. 

In order to determine the dynamic universality class, hydrodynamic variables and conservation laws provide important clues. But they do not determine the dynamic universality class uniquely. {\it It is also important to take the mode-mode coupling strength into account:} this is one lesson of our analysis. We have seen that mode-mode couplings are suppressed in $1/N_c^2$ although they are relevant operators.

There are various attempts to explore $1/N_c$-effects in the AdS/CFT duality. See Ref.~\cite{CaronHuot:2009iq} for long-time power-law tails and Ref.~\cite{Anninos:2010sq} for symmetry restoration. Similarly, it would be interesting to carry out bulk 1-loop computations for dynamic critical phenomena. 

%\clearpage

%%------------------
\subsection{Implications to heavy-ion collisions}\label{sec:heavy_ion}
%%------------------

So far we focused on the ${\cal N}=4$ SYM in the large-$N_c$ limit. In this subsection, we discuss the implication to QCD.

According to lattice results, the finite temperature transition from the hadronic phase to the quark-gluon plasma phase (at zero baryon chemical potential) is not a phase transition but rather a smooth crossover. On the other hand, the finite density transition at $T=0$ is believed to be a first-order phase transition, so there must be an end point of the first-order phase transition line somewhere in the phase diagram, which is a critical point. The precise location of the critical point is still unknown, and one of major goals of future RHIC experiments is this critical end point search. 

In real experiments, singular behaviors at the critical point are limited by the finite size and finite time effects \cite{Berdnikov:1999ph}. The former limits $\xi$ by $\xi < ({\rm size})$ and the typical size of the plasma is 10 fm. However, the latter effect, finite evolution time is more important. Since $\tau \sim \xi^z$, $\xi$ is limited by $\xi < ({\rm time})^{1/z}$. The finite time effect is more important because $z>1$ and the typical evolution time is the same order as the spatial size. 

Son and Stephanov argued that QCD belongs to model H at the critical end point \cite{Son:2004iv}. The baryon number density corresponds to the R-charge density.%
\footnote{Strictly speaking, QCD has an additional hydrodynamic mode compared with model H, which is the QCD chiral condensate. However, Son and Stephanov have shown that only one linear combination of the baryon number density and the chiral condensate becomes truly hydrodynamic. }
For model H, $z \sim 3$, so the finite time effect limits $\xi< 2 \sim 3$ fm compared with the natural value of 1 fm. 

In principle, model B and H behaviors are distinguishable since diffusion constants scale with $\xi$ differently: $D_{\rm H}/D_{\rm B} \propto \xi$. (Here, we focus only on the baryon conductivity or baryon diffusion constant since the shear viscosity has only the mild singularity.) But in real experiments, one does not have enough size and time to see the divergence of $\xi$, and the model H effect is suppressed by $1/N_c^2$. This raises a question whether the model H behavior is distinguishable from the model B one in real experiments. 

In other words, in order to know which effect is dominant, the inspection of $\xi$-dependence is not sufficient.
%It would be useful to estimate not only the $\xi$-dependence but also the dependence on parameters and numerical coefficients. 
Of course, for QCD, $N_c^2$ is not small but not very large either. The other parameters and numerical coefficients would become important to compare model B and H effects. 
Naively, one may imagine that numerical coefficients are not very relevant to the comparison 
%one important dependence is the coupling-constant dependence 
%one is inclined to regard these factors irrelevant for the comparison. 
because the coupling-constant dependence would overwhelm numerical coefficients in the strong coupling limit.
One would expect $D_{\rm B}, \bareta \rightarrow 0$ at strong coupling.%
\footnote{For example, perturbative gauge theories give $D_{\rm B}, \bareta \propto 1/(g_{\rm YM}^4 \ln g_{\rm YM}^{-1})$ at zero chemical potential \cite{Arnold:2000dr}.
%where $g_{\rm YM}$ is the gauge coupling. 
In the naive strong coupling limit $g_{\rm YM} \rightarrow \infty$, they vanish.} 
Since $D_{\rm H} \propto 1/\bareta$, the coupling-constant dependence would magnify the model H effect. However, in the AdS/CFT computations, both $D_{\rm B}$ and $\bareta$ remain finite in the large-$N_c$ limit. 
%Thus, the coupling-constant dependence does not seem to play a very important role here. 
This is the reason why it is necessary to reexamine the issue.

%This requires a serious study of QCD which is beyond the scope of this paper. 
It is beyond the scope of this paper to make a realistic estimate for QCD. Here, we compare model B and H effects using the ${\cal N}=4$ results. 
From Eqs.~(\ref{eq:R-charged_diffusion_H}) and (\ref{eq:R-charged_diffusion_B}), 
\be
\frac{D_{\rm H}}{D_{\rm B}} \sim \frac{256\pi}{9} 0.12 \frac{T_c}{N_c^2} \xi~.
\ee
Thus, the model H effect becomes dominant when
\be
\xi \gg \frac{9}{256\pi \cdot 0.12} \frac{N_c^2}{T_c} \sim \frac{0.75}{T_c}~.
\ee
In the last expression, we used $N_c^2=8$. Thus, for the $SU(3)$ ${\cal N}=4$ SYM, the $1/N_c^2$-suppression does not have a significant effect, and the model H effect becomes compatible with the model B effect as soon as $\xi > 1/T_c$. 

% v2
The above result is valid for the large-$N_c$ ${\cal N}=4$ SYM and not for QCD, but one may be tempted to use the QCD transition temperature for $T_c$. Since $\hbar c \sim 197~{\rm MeV fm}$ and 197 MeV is not far from $T_c$, the characteristic length scale at $T_c$ is $T_c^{-1} \sim 1$ fm.%
\footnote{
Because the ${\cal N}=4$ SYM is scale invariant, only $T_c$ (or $\mu_c$) is the independent scale. This is not the case for QCD, and a realistic estimate would involve $\mu_c$ as well. Also, currently the location of the QCD critical end point is not known very precisely. See, {\it e.g.}, Ref.~\cite{Stephanov:2007fk} for a summary of theoretical results. For simplicity, we use $T_c^{-1} = 1$ fm for the above estimate.
}
Using $T_c^{-1} = 1$ fm, one gets $\xi \gg 0.75$ fm. It would be interesting to make a realistic estimate for QCD. 

%%------------------
\subsection{Static critical exponents of the ${\cal N}=4$ plasma}\label{sec:N=4_static_universality}
%%------------------

The ${\cal N}=4$ SYM has the following static critical exponents in the large-$N_c$ limit:
\be
(\alpha, \beta, \gamma, \delta, \nu, \eta) = \left(\frac{1}{2}, \frac{1}{2}, \frac{1}{2}, 2,  \frac{1}{4}, 0\right)~.
%_{{\cal N}=4}~. 
%_{\rm R-charged}
\label{eq:R-charged_static_exponents}
\ee
In this subsection, we discuss the implications of these exponents. 

Recall some basic facts about the static critical phenomena. 
%\marginpar{?}
In mean-field theories, the statistical fluctuations of the order parameter are ignored. The effect of fluctuations become more important in low spatial dimensions, so mean-field exponents, which do not depend on the dimensionality, can be modified in low dimensions. On the other hand, the effect is less important in high spatial dimensions. So, mean-field exponents are trustable for high enough $d_s>d_c$, where $d_c$ is known as the upper critical dimension. 

One characteristic behavior for $d_s>d_c$ is the breakdown of the hyperscaling relation (\ref{eq:scaling}). The hyperscaling relation often fails above $d_c$. This is because the so-called dangerously irrelevant operators may exist in free energy. In mean-field theories, static exponents are independent of $d_s$, so the hyperscaling relation obviously fails. For mean-field theories, the hyperscaling relation can be satisfied only for $d_s=d_c$. 

Keeping these in mind, let us go back to the ${\cal N}=4$ exponents (\ref{eq:R-charged_static_exponents}). 
First, these values are unconventional compared with the Ginzburg-Landau theory (\ref{eq:static_exponent_GL}). Second, one should regard them as an unconventional but mean-field result since we are taking the large-$N_c$ limit.%
\footnote{In this paper, we draw a distinction between mean-field theory and the Ginzburg-Landau theory. We use ``mean-field theory" when the statistical fluctuations of the order parameter are ignored. The free energy of such a theory may or may not take the Ginzburg-Landau form. %We allow the possibility that the mean-field free energy takes non-Ginzburg-Landau form. 
}
In the large-$N_c$ limit, fluctuations are suppressed so that mean-field results are exact. 
%\marginpar{?}
%\footnote{For the $O(n)$ model, the static exponents in the large-$n$ limit do not reduce to the mean-field results for $d_s<d_c$. This is because we take the large-$N_c$ limit first whereas the large-$n$ limit is taken at the end for the $O(n)$ model.} 
More explicitly, the effect of fluctuations can be estimated from the Ginzburg criterion. Fluctuations are suppressed when
\be
\bra (\rho-\bra\rho\ket)^2 \ket \ll \bra \rho-\rho_c \ket^2~.
%\bra \rho\rho \ket \ll \bra \rho \ket^2~.
%
\ee
Since $\bra (\rho-\bra\rho\ket)^2 \ket = O(N_c^2)$ and $\bra \rho-\rho_c \ket^2 =O(N_c^4)$, the Ginzburg criterion is always satisfied in the large-$N_c$ limit. 
As a mean-field result, one expects that the exponents are independent of $d_s$. For holographic superconductors, this was shown explicitly \cite{Maeda:2009wv}. 
% v2 
The vanishing anomalous exponent $\eta=0$ is another indication of mean-field results since the nonvanishing $\eta$ usually comes from fluctuations.

Third, although the exponents are unconventional, they satisfy the static scaling relations. However, the hyperscaling relation is not satisfied, which is another indication of a mean-field result. 

Now, the story should change at finite $N_c$. Using the static critical exponents, the Ginzburg criterion tells that fluctuations are suppressed for $d>d_c$, where 
\be
d_c = \frac{2\beta+\gamma}{\nu} = \frac{2-\alpha}{\nu}~.
\label{eq:critical_dim}
\ee
%$d_c$ is known as the upper critical dimension. 
%Using the static exponents for the ${\cal N}=4$ plasma (\ref{eq:R-charged_static_exponents}), 
Using the ${\cal N}=4$ exponents, $d_c=6$. In fact, the hyperscaling relation would be satisfied at $d_c$. Thus, the exponents (\ref{eq:R-charged_static_exponents}) should be modified for $d_s=3$ at finite $N_c$. The static exponents should change so that the hyperscaling relation for $d_s=3$ is satisfied, {\it i.e.}, $2-\alpha=3\nu$. 
%\marginpar{?}
It would be interesting to compute $1/N_c$-effects to see the deviation from the ``mean-field" behavior. 
%It is possible that the system approaches to the conventional Ginzburg-Landau theory at finite $N_c$. 

%Also, note that the exponents (\ref{eq:R-charged_static_exponents}) do not satisfy the hyperscaling relations (\ref{eq:scaling}). The hyperscaling relation often fails above $d_c$. This is because the so-called dangerously irrelevant operators may exist in free energy. In mean-field theories, static exponents are independent of $d_s$, so the hyperscaling relation obviously fails. For mean-field theories, the hyperscaling relation can be satisfied only for $d_s=d_c$. In fact, the exponents (\ref{eq:R-charged_static_exponents}) satisfies the hyperscaling relation for $d_s=6$. On the other hand, at finite $N_c$, the static exponents should change so that the hyperscaling relation for $d_s=3$ is satisfied, {\it i.e.}, $2-\alpha=3\nu$.

The R-charged black holes have been constructed for $\ds \neq 3$. Black holes with $\ds =2, 3$, and 5 are particularly important. They have the same static exponents $(\alpha, \beta, \gamma, \delta)$ as in the ${\cal N}=4$ plasma. The exponents $\nu$ and $\eta$ have not been computed for $\ds \neq 3$. Suppose that the anomalous dimension $\eta$ vanishes even for $\ds \neq 3$. The scaling relation then determines $\nu = 1/4$, so $d_c=6$ as well. Then, the static exponents should be modified for those systems as well.

%%%%%%%%%%%%%%%%%%%%%%%%%%%%%%%%%%%%%%%%

\begin{acknowledgments}

We would like to thank Hirotsugu Fujii, Yuki Minami, Kunimasa Miyazaki, Berndt M\"{u}ller, Chiho Nonaka, Masahiro Ohta, Misha Stephanov, and Hirofumi Wada for useful discussions. 
%We would also like to thank Shigeki Sugimoto and Gary Horowitz for their comments on the manuscript.
%MN would also like to thank ``Non-equilibrium quantum field theories and dynamic critical phenomena" at the Yukawa Institute of Theoretical Physics (March 2009) and IPMU Focus Week ``Condensed matter physics meets high energy physics" (Feb.\ 8-12 2010, IPMU)  for their hospitality and for a stimulating environment. 
This research was supported in part by the Grant-in-Aid for Scientific Research (20540285) from the Ministry of Education, Culture, Sports, Science and Technology, Japan.

\end{acknowledgments}

%%%%%%%%%%%%%%%%%%%%%%%%%%%%%%%%%%%%%%%%

\appendix

%%%%%%%%%%%%%%%%%%%%%%%%%%%%%%%%%%%%%%%%
\section{R-charged black hole} \label{sec:R-charged}
%%%%%%%%%%%%%%%%%%%%%%%%%%%%%%%%%%%%%%%%

The ${\cal N}=4$ plasma at a finite chemical potential is dual to the five-dimensional Einstein-Maxwell-scalar system%
\footnote{
Capital Latin indices $M, N, \ldots$ run through bulk spacetime coordinates $(t, \bmx, u)$, where $(t,\bmx)=(t,x_i)$ are the boundary coordinates and $u$ is the AdS radial coordinate. Greek indices $\mu, \nu, \ldots$ run though only the boundary coordinates. }:
\bea
S_5 &=& \frac{1}{16\pi G_5} \int d^5 x \sqrt{-g}\, 
   \left[ R - \frac{L^2}{8}\, H^{4/3}\, F_{MN}^2 \right.
\nonumber \\
  && \left. - \frac{1}{3} \frac{(\nabla_M H)^2}{H^2}
  + 2 {\cal V} \right]~,
\label{eq:reduced-action-1charge}
\eea
where 
\be
2{\cal V} = \frac{4}{L^2}\, \left( H^{2/3} + 2 H^{-1/3} \right)~,
\ee
and $L$ is the AdS radius. The solution is known as the (single) R-charge black hole \cite{Behrndt:1998jd,Kraus:1998hv,Cvetic:1999xp}:
\begin{subequations}
\bea
ds_5^2
  &=& \frac{(\pi\, T_0\, L)^2}{u}\, H^{1/3}\,
    \left( - \frac{f}{H}\, dt^2 + d\bmx^2 \right) \nonumber \\
%  &=& \frac{(\pi\, T_0\, L)^2}{u}\, H^{1/3}\,
%    \left( - \frac{f}{H}\, dt^2 + dx^2 + dy^2 + dz^2 \right) \nonumber \\
  &&+ \frac{L^2}{4\, f\, u^2}\, H^{1/3}~du^2~,
\label{eq:STU_metric-five_dim-1charge} \\
  A_M
  &=& \pi\, T_0\, \sqrt{2\, \kappa\, (1+\kappa)}~\frac{u}{H}~( dt )_M~,
\label{eq:STU_U1-1charge} \\
  H &=& 1 + \kappa\, u~,
\\
  f &=& H \! - \! (1\! +\! \kappa) u^2
  = (1\! -\! u) \left\{ 1\!  + \! (1\! +\! \kappa) u \right\}~,
\label{eq:STU_scalar-1charge}
\eea
\end{subequations}
where $u \in [0,1]$, with $u=1$ corresponding to the location of the horizon. The parameter $\kappa$ is related to the R-charge (see below), and $T_0$ is the Hawking temperature for the neutral black hole ($\kappa=0$). Thermodynamic quantities are given by
\begin{subequations}
\label{eq:thermodynamic_quantities}
\bea
\epsilon &=& \frac{3\, \pi^2\, N_c^2\, T_0^4}{8}~( 1 + \kappa )~, \\
P &=& \frac{\pi^2\, N_c^2\, T_0^4}{8}~( 1 + \kappa )~, \\
s &=& \frac{\pi^2\,N_c^2\,T_0^3}{2}~\sqrt{1 + \kappa}~,\\
T &=& \frac{2 + \kappa}{2\, \sqrt{1 + \kappa}}~T_0~, 
\label{eq:temperature} \\
\rho &=& \frac{\pi\, N_c^2\, T_0^3}{8}~\sqrt{2\, \kappa\, ( 1 + \kappa )}~,
\label{eq:charge_density} \\
\mu &=& \pi\, T_0\, \sqrt{ \frac{2\, \kappa}{1 + \kappa} }~.
\eea
\end{subequations}
Note that $T$ and $\mu$ satisfy
\be
\frac{2\, \pi\, T}{\mu}
  = \sqrt{ \frac{\kappa}{2} } + \sqrt{ \frac{2}{\kappa} } \ge 2~.
\label{eq:sol-degenerateI}
\ee
Equation~(\ref{eq:sol-degenerateI}) is invariant under $\kappa \rightarrow 4/\kappa$. This suggests that there are two values of $\kappa$ for a given pair of $(T, \mu)$. One can show that the solutions with $\kappa \leq 2$ has the lower free energy, so we focus on the branch.

These thermodynamic quantities lead to the R-charge susceptibility and the specific heat at constant $\mu$:
\bea
   \chi
  &:=& \left( \frac{\partial \rho}{\partial \mu} \right)_{T}
  = \frac{N_c^2 T_0^2}{8}~
     \frac{2 + 5 \kappa - \kappa^2}{2 - \kappa}~,
\label{eq:electric_susceptibility-1charge} \\
   C_\mu
  &:=& T \left( \frac{\partial s}{\partial T} \right)_{\mu}
  = \frac{1}{2} \pi^2 N_c^2 T_0^3~
     \frac{ ( 3 - \kappa ) ( 2 + \kappa ) }
          { \sqrt{1 + \kappa } ( 2 - \kappa ) }~.
\label{eq:heat_capacity-grand_canonical-1charge}
\eea
They diverge at $\kappa = 2$ corresponding to $\pi T/\mu=1$, where the system undergoes a second-order phase transition. 

Using the conventional definition of static critical exponents, the system has
\be
(\alpha, \beta, \gamma, \delta, \nu, \eta) = \left(\frac{1}{2}, \frac{1}{2}, \frac{1}{2}, 2,  \frac{1}{4}, 0\right)~.
\ee
The evaluation of $\nu$ and $\eta$ requires the correlation function, which has been computed in Ref.~\cite{Buchel:2010gd}.

%%%%%%%%%%%%%%%%%%%%%%%%%%%%%%%%%%%%%%%%
\section{Mode-mode coupling computation of model H} \label{sec:H_mode_coupling}
%%%%%%%%%%%%%%%%%%%%%%%%%%%%%%%%%%%%%%%%

In this appendix, we present the mode-mode coupling computation for relativistic fluids. Our computation is similar to the one in Refs.~\cite{kawasaki_shear,ferrell_shear}.

In hydrodynamics, the spatial parts of conserved currents are not independent variables; rather they are determined from conserved charges via the constitutive equations. For the linear deviation from the equilibrium, the constitutive equation for the charge current is given by
\be
J_i(t,\bmx) = - D \del_i\rho(t,\bmx)~.
\ee
At quadratic order, one can add a nonlinear term: 
\be
J_i(t,\bmx) = - D \del_i\rho(t,\bmx) + \frac{1}{\barw} \rho(t,\bmx)  \pi_i(t,\bmx)~,
\label{eq:constitutive_nonlinear}
\ee
where
\bea
 \pi_i(t,\bmx) &:=& T^0_{~i}(t,\bmx)~, \\
\barw &:=& \bar{\epsilon} + \bar{P}~.
\eea
The quantities with \={ } are the values at a chosen equilibrium state. The nonlinear term of \eq{constitutive_nonlinear} represents the effect of convection and gives strong enhancement of conductivity near the critical point. 

The conductivity can be derived from the Kubo formula:
%\marginpar{?}
\bea
\lambda_R(q)  
&=& \frac{1}{\ds T} \! \int_0^\infty \! dt \! \int \! d\bmx\, e^{-i \bmq \cdot \bmx} 
   \bra \bmJ(t,\bmx) \cdot \bmJ(0,\bmo)  \ket \\
&\simeq&  \lambda_B + \frac{1}{\ds \barw^2 T} \int_0^\infty dt \int d\bmx\, e^{-i \bmq \cdot \bmx} 
\nonumber \\
&&  \hspace{0.2in} \times \bra \rho(t,\bmx) \bmPi(t,\bmx) \cdot \rho(0,\bmo) \bmPi(0,\bmo)  \ket~.
\label{eq:method1}
\eea
% correlator vs Green fn
%Here, we assume that the convection term becomes dominant near the critical point.
The mode-mode coupling theory evaluates the four-point correlator as follows. First, approximate the four-point correlator into the product of the two-point correlators as
\bea
\Delta \lambda(q) &:=& \lambda_R(q) - \lambda_B 
\nonumber \\
&\simeq& \frac{1}{\ds \barw^2 T} \int_0^\infty dt \int d\bmx\, e^{-i \bmq \cdot \bmx} 
\nonumber \\
&&  \hspace{0.2in}\times \bra \rho(t,\bmx) \rho(0,\bmo) \ket
   \bra \bmPi(t,\bmx) \cdot \bmPi(0,\bmo)  \ket~.
\eea
We will use the Fourier-transformed correlators
\bea
\bra \rho(t,\bmx) \rho(0,\bmo) \ket &=& 
\! \int \! \frac{d\bmp}{(2\pi)^{\ds}}\,e^{i \bmp \cdot \bmx} \tilG_{\rho}(t,\bmp)~, \\
\frac{1}{2} \bra \{ \pi_{i}(t,\bmx), \pi_{j}(0,\bmo) \} \ket &=&
\! \int \! \frac{d\bmp}{(2\pi)^{\ds}}\,e^{i \bmp \cdot \bmx} \tilG_{ij}(t,\bmp)~.
\eea
(The symmetrized correlator $ \bra \{ \pi_{i}(x), \pi_{j}(0) \} \ket/2 $ and the Wightman correlator $ \bra \pi_{i}(x) \pi_{j}(0) \ket $ make no difference for classical fields.)
Second, use the results of the linearized theory for the correlators:
\bea
\tilG_\rho(t,\bmp) &\simeq& e^{-Dp^2|t|} T\chi_p
:= e^{-Dp^2|t|} \frac{T \chi}{1+p^2 \xi^2}~, \\
\tilG_{ij}(t,\bmp) &\simeq& \barw T 
\left\{ (\delta_{ij} - \hp_i \hp_j) e^{ -\gamma_\eta p^2 |t| } \right. 
\nonumber \\
&& + \left. \hp_i \hp_j e^{ -\gamma_s p^2 |t|/2 } \cos(p c_s t) \right\}~,
\label{eq:correlation_mom}
\eea
%where $\bareta$ is the renormalized shear viscosity. 
where $\hat{\bmp} = \bmp/p$ and
\bea
\gamma_\eta &:=& \frac{\bareta}{\barw}~, \\
\gamma_s &:=& \frac{\zeta+2\frac{\ds-1}{\ds} \bareta}{\barw}~.
\eea
The first and second terms of \eq{correlation_mom} come from the transverse shear modes and the longitudinal sound mode, respectively. Finally, the timescale of $\rho$ is much slower than that of $\bmPi$, so we will set $D=0$.
Then, $(\bmp:=\bmq-\bmk)$
%space -> \!
\begin{eqnarray*}
\lefteqn{ \Delta\lambda(q) }\nonumber \\
&&\simeq \frac{1}{\ds \barw^2 T} \int_0^\infty dt \int \frac{d\bmk}{(2\pi)^{\ds}}\,
\tilG_{\rho} (t,\bmk) \delta^{ij} \tilG_{ij} (t,\bmp)
%\qquad (\bmp:=\bmq-\bmk) 
\\
&&\simeq \frac{T}{\ds \barw} \int_0^\infty dt \int \frac{d\bmk}{(2\pi)^{\ds}}\,
\chi_k \left[ (\ds-1) e^{ -\gamma_\eta p^2 |t| } \right.
\nonumber \\
&& \hspace{0.5in} 
\left. + e^{ -\gamma_s p^2 |t|/2 } \cos(p c_s t) \right] \\
&&= \frac{T}{\ds \barw} \int \frac{d\bmk}{(2\pi)^{\ds}}\,
\chi_k \left[ \frac{\ds-1}{\gamma_\eta} \frac{1}{p^2}
+ \frac{2}{\gamma_s} \frac{1}{p^2+(2c_s/\gamma_s)^2} \right]~.
\end{eqnarray*}
 
We henceforth perform the integration for $\ds=3$ and $q=0$. One obtains
\be
\Delta\lambda
= \frac{T \chi}{6\pi (\barw \gamma_\eta)} \frac{1}{\xi}
+ \frac{T \chi}{6\pi (\barw \gamma_s)} \frac{1}{\xi [1+(2c_s\xi/\gamma_s)]}~.
\ee
As long as $c_s \xi/\gamma_s \gg 1$, the longitudinal sound mode is negligible as expected, and only the transverse shear modes contribute to $\Delta\lambda$:
\be
\lambda_R = \lambda_{\rm B} + \frac{T}{6\pi\bareta} \chi \frac{1}{\xi}~.
\ee

A similar computation can be done for $q\neq0$. The transverse shear part gives
\be
\Delta\lambda(q)
= \frac{T \chi_q}{6\pi (\barw \gamma_\eta)} \frac{1}{\xi} K(q\xi)~.
\ee
where
\be
K(x) := \frac{3}{4x^2} \{ 1+ x^2+(x^3 - x^{-1}) \tan^{-1} x \}~,
\ee
which behaves as
\be
K(x) \sim \left\{
   \begin{array}{ll}
	1 & (x \ll 1) \\
	\frac{3\pi}{8} x & (x \gg 1)~.
   \end{array}
\right.
\ee
The function $x^2 K(x)$ is the so-called Kawasaki function $K_0(x)$. 
Thus, the dispersion relation $\omega = -i\lambda \chi_q^{-1} q^2$ becomes
%\be
%
%\omega \sim \left\{
%   \begin{array}{ll}
%   	-i \frac{T}{6\pi\bareta}\frac{1}{\xi} q^2 & (q\xi \ll 1) \\
%	-i \frac{T}{16\pi\bareta} q^3 & (q\xi \gg 1)~.
%   \end{array}
%\right.
%\label{eq:scaling_ex}
%
%\ee
\bea
\omega 
&\sim& -i \frac{T}{6\pi\bareta}\frac{1}{\xi} q^2 
\quad (q\xi \ll 1)~,
\label{eq:hydro_regime} \\
&\sim& -i \frac{T}{16\bareta} q^3 
\quad (q\xi \gg 1)~.
\label{eq:critical_regime}
\eea
Equations~(\ref{eq:hydro_regime}) and (\ref{eq:critical_regime}) are an explicit realization of the scaling argument in \sect{A&B}. In the hydrodynamic regime $q\xi \ll 1$, $\omega \propto q^2/\xi \sim (q\xi)^2/\xi^3$. On the other hand, in the critical regime $q\xi \gg 1$, $\omega \propto q^3 \sim (q\xi)^3/\xi^3$. They match smoothly at $q\xi \sim 1$.

A similar computation of the energy-momentum tensor correlator determines $\bareta_R$ and its exponent $x_\eta$ (\ref{eq:x_eta-def}), from which one can determine the exponent $x_\lambda$.


\begin{thebibliography}{}


\bibitem{textbook}
  J. Cardy,
  {\it Scaling and renormalization in statistical physics}
  (Cambridge Univ.\ Press, Cambridge, 1996).
%  W. Gebhardt and U. Krey,
%  {\it Phasen\"{u}berg\"{a}nge und kritische Ph\"{a}nomene}
%  (Vieweg \& Sohn, Braunschweig/Wiesbaden, 1980); 
%  A.~Onuki, {\it Phase transition dynamics} 
%  (Cambridge Univ.\ Press, Cambridge, 2002).

%\bibitem{vasilev}
%A.N.~Vasil'ev, {\it The field theoretic renormalization group in critical behavior theory and stochastic dynamics} (CRC Press LLC, Boca Raton, Florida, 2004) Sect.~5.24.

%%% AdS/CFT %%%

%\cite{Maldacena:1997re}
\bibitem{Maldacena:1997re}
  J.~M.~Maldacena,
``The large N limit of superconformal field theories and supergravity,''
  Adv.\ Theor.\ Math.\ Phys.\  {\bf 2} (1998) 231
  [Int.\ J.\ Theor.\ Phys.\  {\bf 38} (1999) 1113]
  [arXiv:hep-th/9711200].
  %%CITATION = HEP-TH 9711200;%%

%\cite{Witten:1998qj}
\bibitem{Witten:1998qj}
E.~Witten,
``Anti-de Sitter space and holography,''
Adv.\ Theor.\ Math.\ Phys.\  {\bf 2} (1998) 253
[arXiv:hep-th/9802150].
%%CITATION = HEP-TH 9802150;%%

%\cite{Witten:1998zw}
\bibitem{Witten:1998zw}
E.~Witten,
``Anti-de Sitter space, thermal phase transition, and confinement in  gauge theories,''
Adv.\ Theor.\ Math.\ Phys.\  {\bf 2} (1998) 505 
[arXiv:hep-th/9803131].
%%CITATION = HEP-TH 9803131;%%

%\cite{Gubser:1998bc}
\bibitem{Gubser:1998bc}
  S.~S.~Gubser, I.~R.~Klebanov and A.~M.~Polyakov,
  ``Gauge theory correlators from non-critical string theory,''
  Phys.\ Lett.\ B {\bf 428} (1998) 105
  [arXiv:hep-th/9802109].
  %%CITATION = HEP-TH 9802109;%%
  
%%% AdS/QGP %%%

%\cite{Natsuume:2007qq}
%\bibitem{Natsuume:2007qq}
%  M. Natsuume,
%  ``String theory and quark-gluon plasma,''
%  arXiv:hep-ph/0701201.
  %%CITATION = HEP-PH/0701201;%%

%\cite{Natsuume:2008ha}
%\bibitem{Natsuume:2008ha}
%  M.~Natsuume,
%  ``String theory implications on causal hydrodynamics,''
%  Prog.\ Theor.\ Phys.\ Suppl.\  {\bf 174} (2008) 286
%  [arXiv:0807.1394 [nucl-th]].
  %%CITATION = PTPSA,174,286;%%

%%% Dynamic critical phenomena %%%

%\cite{Maeda:2008hn}
\bibitem{Maeda:2008hn}
  K.~Maeda, M.~Natsuume and T.~Okamura,
  ``Dynamic critical phenomena in the AdS/CFT duality,''
  Phys.\ Rev.\  D {\bf 78} (2008) 106007
  [arXiv:0809.4074 [hep-th]].
  %%CITATION = PHRVA,D78,106007;%%
  
%\cite{Maeda:2009wv}
\bibitem{Maeda:2009wv}
  K.~Maeda, M.~Natsuume and T.~Okamura,
  ``Universality class of holographic superconductors,''
  Phys.\ Rev.\  D {\bf 79} (2009) 126004
  [arXiv:0904.1914 [hep-th]].
  %%CITATION = PHRVA,D79,126004;%%

%\cite{Buchel:2009mf}
\bibitem{Buchel:2009mf}
  A.~Buchel and C.~Pagnutti,
  ``Transport at criticality,''
  Nucl.\ Phys.\  B {\bf 834} (2010) 222
  [arXiv:0912.3212 [hep-th]].
  %%CITATION = NUPHA,B834,222;%%

%\cite{Buchel:2010gd}
\bibitem{Buchel:2010gd}
  A.~Buchel,
  ``Critical phenomena in N=4 SYM plasma,''
  arXiv:1005.0819 [hep-th].
  %%CITATION = ARXIV:1005.0819;%%

%\cite{Natsuume:2010vb}
\bibitem{Natsuume:2010vb}
  M.~Natsuume,
  ``Critical phenomena in the AdS/CFT duality,''
  arXiv:1006.4930 [hep-th].
  %%CITATION = ARXIV:1006.4930;%%

%\cite{Buchel:2010ys}
\bibitem{Buchel:2010ys}
  A.~Buchel and C.~Pagnutti,
  ``Critical phenomena in N=2* plasma,''
  arXiv:1010.3359 [hep-th].
  %%CITATION = ARXIV:1010.3359;%%

%%%

\bibitem{hohenberg_halperin}
  P.C. Hohenberg and B.I. Halperin,
``Theory of dynamic critical phenomena,"
  Rev.\ Mod.\ Phys.\  {\bf 49} (1977) 435.
  
%%% R-charged black hole %%%

%\cite{Behrndt:1998jd}
\bibitem{Behrndt:1998jd}
  K. Behrndt, M. Cvetic and W.A. Sabra,
  ``Non-extreme black holes of five dimensional N = 2 AdS supergravity,''
  Nucl.\ Phys.\ B {\bf 553} (1999) 317
  [arXiv:hep-th/9810227].
  %%CITATION = HEP-TH 9810227;%%

%\cite{Kraus:1998hv}
\bibitem{Kraus:1998hv}
  P. Kraus, F. Larsen and S.P. Trivedi,
  ``The Coulomb branch of gauge theory from rotating branes,''
  JHEP {\bf 9903} (1999) 003
  [arXiv:hep-th/9811120].
  %%CITATION = HEP-TH 9811120;%%

%\cite{Cvetic:1999xp}
\bibitem{Cvetic:1999xp}
  M. Cvetic {\it et al.},
  ``Embedding AdS black holes in ten and eleven dimensions,''
  Nucl.\ Phys.\ B {\bf 558} (1999) 96
  [arXiv:hep-th/9903214].
  %%CITATION = HEP-TH 9903214;%%

%%%

%\cite{Son:2004iv}
\bibitem{Son:2004iv}
  D.~T.~Son and M.~A.~Stephanov,
  ``Dynamic universality class of the QCD critical point,''
  Phys.\ Rev.\  D {\bf 70} (2004) 056001
  [arXiv:hep-ph/0401052].
  %%CITATION = PHRVA,D70,056001;%%

%%% R-charged black hole (static universality) %%%

%\cite{Gubser:1998jb}
\bibitem{Gubser:1998jb}
  S.~S.~Gubser,
  ``Thermodynamics of spinning D3-branes,''
  Nucl.\ Phys.\  B {\bf 551} (1999) 667
  [arXiv:hep-th/9810225].
  %%CITATION = NUPHA,B551,667;%%

%\cite{Cai:1998ji}
\bibitem{Cai:1998ji}
  R.~G.~Cai and K.~S.~Soh,
  ``Critical behavior in the rotating D-branes,''
  Mod.\ Phys.\ Lett.\  A {\bf 14} (1999) 1895
  [arXiv:hep-th/9812121].
  %%CITATION = MPLAE,A14,1895;%%

%\cite{Cvetic:1999rb}
\bibitem{Cvetic:1999rb}
  M.~Cvetic and S.~S.~Gubser,
  ``Thermodynamic stability and phases of general spinning branes,''
  JHEP {\bf 9907} (1999) 010
  [arXiv:hep-th/9903132].
  %%CITATION = JHEPA,9907,010;%%

%%% R-charged BH (shear viscosity) %%%

%\cite{Mas:2006dy}
\bibitem{Mas:2006dy}
  J.~Mas,
  ``Shear viscosity from R-charged AdS black holes,''
  JHEP {\bf 0603} (2006) 016
  [arXiv:hep-th/0601144].
  %%CITATION = HEP-TH 0601144;%%

%\cite{Son:2006em}
\bibitem{Son:2006em}
  D.~T.~Son and A.~O.~Starinets,
  ``Hydrodynamics of R-charged black holes,''
  JHEP {\bf 0603} (2006) 052
  [arXiv:hep-th/0601157].
  %%CITATION = HEP-TH 0601157;%%

%\cite{Saremi:2006ep}
\bibitem{Saremi:2006ep}
  O.~Saremi,
  ``The viscosity bound conjecture and hydrodynamics of M2-brane theory at finite chemical potential,''
  JHEP {\bf 0610} (2006) 083
  [arXiv:hep-th/0601159].
  %%CITATION = HEP-TH 0601159;%%

%\cite{Maeda:2006by}
\bibitem{Maeda:2006by}
  K.~Maeda, M.~Natsuume and T.~Okamura,
  ``Viscosity of gauge theory plasma with a chemical potential from AdS/CFT correspondence,''
  Phys.\ Rev.\ D {\bf 73} (2006) 066013
  [arXiv:hep-th/0602010].
  %%CITATION = HEP-TH 0602010;%%

%%%

\bibitem{kawasaki_bulk}
K.~Kawasaki, 
``Sound attenuation and dispersion near the liquid-gas critical point,"
Phys.\ Rev.\ A {\bf 1} (1970)1750.

\bibitem{onuki_bulk}
A.~Onuki, 
``Dynamic equations and bulk viscosity near the gas-liquid critical point,"
Phys.\ Rev.\ E {\bf 55} (1997) 403.

%%% 1/N %%%%

%\cite{Kovtun:2003vj}
\bibitem{Kovtun:2003vj}
  P.~Kovtun and L.~G.~Yaffe,
  ``Hydrodynamic fluctuations, long-time tails, and supersymmetry,''
  Phys.\ Rev.\  D {\bf 68} (2003) 025007
  [arXiv:hep-th/0303010].
  %%CITATION = PHRVA,D68,025007;%%

%\cite{CaronHuot:2009iq}
\bibitem{CaronHuot:2009iq}
  S.~Caron-Huot and O.~Saremi,
  ``Hydrodynamic Long-Time tails From Anti de Sitter Space,''
  arXiv:0909.4525 [hep-th].
  %%CITATION = ARXIV:0909.4525;%%

%\cite{Anninos:2010sq}
\bibitem{Anninos:2010sq}
  D.~Anninos, S.~A.~Hartnoll and N.~Iqbal,
  ``Holography and the Coleman-Mermin-Wagner theorem,''
  arXiv:1005.1973 [hep-th].
  %%CITATION = ARXIV:1005.1973;%%

%%%

%\cite{Berdnikov:1999ph}
\bibitem{Berdnikov:1999ph}
  B.~Berdnikov and K.~Rajagopal,
  ``Slowing out of equilibrium near the QCD critical point,''
  Phys.\ Rev.\  D {\bf 61} (2000) 105017
  [arXiv:hep-ph/9912274].
  %%CITATION = PHRVA,D61,105017;%%

\bibitem{Arnold:2000dr}
  P.~B.~Arnold, G.~D.~Moore and L.~G.~Yaffe,
  ``Transport coefficients in high temperature gauge theories: (I) Leading-log
  results,''
  JHEP {\bf 0011} (2000) 001
  [arXiv:hep-ph/0010177].
  %%CITATION = JHEPA,0011,001;%%

%\cite{Stephanov:2007fk}
\bibitem{Stephanov:2007fk}
  M.~A.~Stephanov,
  ``QCD phase diagram: An Overview,''
  PoS {\bf LAT2006 } (2006)  024.
  [hep-lat/0701002].
  
%%% mode coupling %%%

\bibitem{kawasaki_shear}
K.~Kawasaki, 
``Kinetic equations and time correlation functions of critical fluctuations," Annals Phys.\ {\bf 61} (1970) 1.

\bibitem{ferrell_shear}
R.~A.~Ferrell, 
``Decoupled-mode dynamical scaling theory of the binary-liquid phase transition,"
Phys.\ Rev.\ Lett.\ {\bf 24} (1970) 1169.

\end{thebibliography}
\end{document}